\documentclass[letterpaper,twocolumn,superscriptaddress,floatfix]{revtex4}
\usepackage[utf8]{inputenc}
\usepackage{graphicx}
\usepackage{subfigure}
\usepackage{color}
\usepackage{amsmath}
\usepackage[margin=0.5in]{geometry}
\usepackage{ulem}
\newcommand{\ket}[1]{|#1\rangle} 
\newcommand{\bra}[1]{\langle#1|} 

\begin{document}
\title{Transverse Confinement of Photon Position in Light-Atom Interaction}

\author{Jun Sun$\footnote{email:sunjun22@mail.ustc.edu.cn}$}
\affiliation{Wuhan Institute of Physics and Mathematics, Chinese Academy of Sciences, Wuhan 430071, China}

\author{Yong-Nan Sun}
\affiliation{CAS Key Laboratory of Quantum Information, University of Science and Technology of China, Hefei, 230026, China}
\affiliation{CAS Center For Excellence in Quantum Information and Quantum Physics, University of Science and Technology of China, Hefei, 230026, China}

\begin{abstract}
{In light-pulsed atom interferometry, the phase accumulated by atoms depends on the effective wave vector of the absorbed photons. In this work, we proposed a theory model to analyses the effective wave vector of photons in structured light. As for monochromatic optical field, a transverse confinement could lead to diffraction. We put forward that in light-atom interaction, the atom wave function could also provide a transverse confinement thus affect the effective wave vector of the absorbed photons. We calculated the relative shift of the photon effective wave vector when an atom with a Gaussian wave function absorbs one photon at the waist in a Gaussian beam. This shift could lead to a systematic effect related to atom spatial distribution in high precision experiment based on light-pulsed atom interferometry.}
\end{abstract}


\maketitle
\textit{Introduction.---}In quantum optics, structured light field could be decomposed into a superposition of plane waves. When an atom absorbs one photon in a plane wave, the transferred momentum is supposed to be along with the direction of the plane wave propagation.

In light-pulsed atom interferometry, photons transfer momentum to atoms to split and interfere matter waves \cite{RN5}. Those interferometers are applied as high precision inertial sensors to measure gravitational acceleration \cite{RN3,RN6}, gravity gradient \cite{RN7,RN8}, fine structure constant \cite{RN9,RN12} and Newton's gravitational constant \cite{RN13,RN1}. The phase shift accumulated by atoms is dependent on the effective wave vector $\vec{k}_{eff}$ of the absorbed photons \cite{RN10}. The deviation of $\vec{k}_{eff}$ in a structured optical field from the wave vector in plane wave light could lead to the systematic effect of ``wavefront curvature".

At present in atom interferometry, $\vec{k}_{eff}$ is supposed to be the local wave vector, give by the phase gradient $\vec{\nabla}\phi$ \cite{RN2} at the atom position and the transferred momentum is supposed to be the canonical photon momentum $\vec{p}_{can}=\hbar\vec{\nabla}\phi$ \cite{RN11,RN14, RN15}. Here $\phi$ is the phase of the optical field.

However there are two problems with this approach. First, the canonical photon momentum doesn't comply with the uncertainty principle, as by making the phase gradient one may define the position and momentum of photon at the same time. Second, in structured monochromatic field, such as Gaussian beam, at some region the local phase gradient is larger than the wavenumber $k$. If one atom absorbs one photon at this region, the transferred canonical momentum could be larger than $\hbar k$ \cite{RN11,RN15}. While if we decompose light into plane waves, in each mode the photon momentum should not be larger than $\hbar k$ along any direction. 

To solve those two problems, in this work based on Fourier optics \cite{RN16}, we provide a theory model to estimate $\vec{k}_{eff}$ of photons in structured light field. As transverse confinement could lead to diffraction, we proposed that when an atom absorbs one photon, the atom position wave function could also provide a transverse confinement to light thus affect $\vec{k}_{eff}$ of the absorbed photons. The relative shift of effective wave vector could lead to a systematic effect related to atom spacial distribution in light-pulsed atom interferometry. 

Our model could provide an alternative interpretation for the extra photon recoil observed in a distorted optical field \cite{RN11,RN14}.

\begin{figure*}[t]
	\centering
	\includegraphics[width=0.95\linewidth]{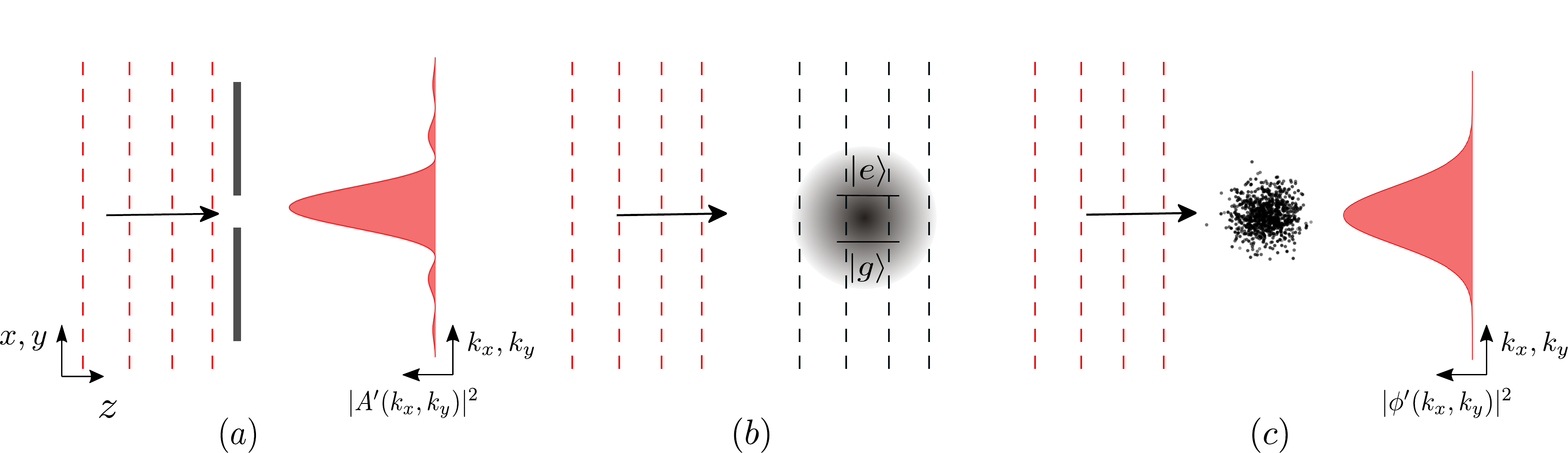}
	\caption{(a) An incident plane wave (red dashed line), $u(x,y,z) = e^{ikz}$ diffracted by a single slit. The transmission rate of the slit is $t(x,y)=rect(x/a)$ and $a$ is the slit width. The modulus square of the angular spectrum of the diffracted light is $|A'(k_x,k_y)|^2=a^2sinc(ak_x/2\pi)^2$. (b) Geometric illustration of the interaction between a plane wave light field and a plane atomic matter wave (black dashed line). The atom wave function is $\psi_{atom}=e^{ipz/\hbar}$. There is no transverse confinement in this case. (c) Interaction between a plane wave light and a 3D Gaussian atomic cloud (black dots) with radius $\sigma_a$. The atom transverse probability distribution is $|\psi_{atom}(x,y)|^2 = (1/2\pi\sigma_a^2)e^{-i(x^2+y^2)/4\sigma_a^2}$. The atom position distribution could provide a transverse confinement to the absorbed photons and lead to a projection of their transverse position state. For the absorbed photons, the transverse wave vector probability distribution becomes $|\phi'(k_x,k_y)|^2=(1/\pi\sigma_a^2)e^{-\sigma_a^2(k_x^2+k_y^2)}$.}
	\label{Fig.1}
\end{figure*}

\textit{Effective wave vector.}---We start with a monochromatic scalar optical field with complex amplitude $\Psi(x,y,z,t) = u(x,y,z)e^{-i\omega t}$. In vacuum $u(x,y,z)$ applies for the Helmholtz equation:
\begin{equation}
	(\nabla^2 + k^2)u(x,y,z) = 0
\end{equation}
$k=\omega/c$. Here $\omega$ is angular frequency and $c$ is the speed of light. Structured field could be decomposed into a superposition of plane waves with angular spectrum $A(k_x,k_y)$ \cite{RN17}
\begin{equation}
\label{E2}
\begin{aligned}
u(x,y,z)= \frac{1}{4\pi^2}\iint &A(k_x,k_y)e^{i(k_xx+k_yy)}\\
&\times e^{i\sqrt{k^2-k_x^2-k_y^2}z}dk_xdk_y.
\end{aligned}
\end{equation}
$A(k_x,k_y)$ is the Fourier transform of $u(x,y,0)$, whose modulus square describes the weighting of each plane wave mode
\begin{equation}
\label{E3}
A(k_x,k_y) =\iint u(x,y,0)e^{-i(k_xx+k_yy)}dxdy.
\end{equation}

To calculate $\vec{k}_{eff}$ we normalize the integration of $|A(k_x,k_y)|^2$ and define two relative complex amplitude functions:
\begin{equation}
\phi(k_x,k_y) = \frac{ A(k_x,k_y)}{\sqrt{\iint |A(k_x,k_y)|^2 dk_xdk_y}},
\end{equation}
and
\begin{equation}
\label{E5}
\begin{aligned}
\psi(x,y,z)= \frac{1}{4\pi^2}\iint &\phi(k_x,k_y)e^{i(k_xx+k_yy)}\\
&\times e^{i\sqrt{k^2-k_x^2-k_y^2}z}dk_xdk_y.
\end{aligned}
\end{equation}
$\psi(x,y,z_0)$ is the photon transverse position probability amplitude on plane $z=z_0$, and $\phi(k_x,k_y)$ is the photon transverse wave vector probability amplitude. We have the photon transverse state on plane $z_0$ as:
\begin{equation}
\label{E6}
\begin{aligned}
\ket{photon} &= \iint\psi(x,y,z_0)\ket{x,y}dxdy\\
&=\iint \phi(k_x,k_y)\ket{k_x,k_y}dk_xdk_y.
\end{aligned}
\end{equation}  
$\ket{k_x,k_y}$ represents for plane wave basis $e^{-i(k_xx+k_yy+\sqrt{k^2-k_x^2-k_y^2}z)}$.

Associating with the momentum operator $\hat{p} = -i\hbar\vec{\nabla}$, the effective photon momentum in the field is
\begin{equation}
\label{E7}
\vec{p}_{eff} = \bra{photon}\hat{p}\ket{photon} =  \hbar\vec{k}_{eff} ,
\end{equation}
and the effective wave vector is
\begin{equation}
\label{E8}
\begin{aligned}
\vec{k}_{eff} = &\iint (k_x\vec{e}_{k_x}+k_y\vec{e}_{k_y}\\
    &+\sqrt{k^2-k_x^2-k_y^2}\vec{e}_{k_z})|\phi(k_x,k_y)|^2dk_xdk_y.
\end{aligned}
\end{equation}

\textit{Transverse confinement of light.}---Suppose light propagates towards $+z$ direction. As in monochromatic field $k_x^2 + k_y^2 + k_z^2 = k^2$, $u(x,y,z)$ has only two degrees of freedom. A transverse confinement on plane $z_0$ will affect the field and $\vec{k}_{eff}$ after the plane.  Consider an aperture with transmission rate $t(x,y)$ on the plane 
\begin{equation}
\label{E9}
u'(x,y,z_0) = u(x,y,z_0)t(x,y).
\end{equation}
According to Eq. (\ref{E2}) after $z_0$ the angular spectrum becomes
\begin{equation}
\label{E10}
A'(k_x,k_y) = \mathcal{F}\{u'(x,y,z_0)\}e^{-i\sqrt{k^2-k_x^2-k_y^2}z_0}.
\end{equation}
For $z>z_0$
\begin{equation}
u'(x,y,z) = \mathcal{F}^{-1}\{A'(k_x,k_y)e^{i\sqrt{k^2-k_x^2-k_y^2}z}\}.
\end{equation}
$\mathcal{F}$ and $\mathcal{F}^{-1}$ are Fourier and inverse Fourier transforms. 

Fig. \ref{Fig.1}(a) shows the transverse confinement from a single slit, $t(x,y) = rect(x/a)$, to a plane wave light field $u(x,y,z) = e^{ikz}$. $a$ is the slit width. According to Eq. (\ref{E9}) and Eq. (\ref{E10}), we get the modulus square of the angular spectrum for the diffracted field as $|A'(k_x,k_y)|^2=a^2sinc(ak_x/2\pi)^2$. 

In the point of view of photons, the aperture projects the photon transverse position state and affects its momentum distribution as a result.   

We design a projection operator $\hat{O}$ that projects the photon state to $\ket{x,y}$ with probability $t(x,y)$
\begin{equation}
\label{E12}
\hat{O} = \iint t(x,y)\ket{x,y}\bra{x,y}dxdy.
\end{equation}
After projection, the photon state becomes
\begin{equation}
\ket{photon'} = \hat{O}\ket{photon}.
\end{equation}
On plane $z_0$
\begin{equation}
\label{E14}
\psi'(x,y,z_0) =\frac{\psi(x,y,z_0)t(x,y)}{\sqrt{\iint |\psi(x,y,z_0)t(x,y)|^2 dxdy}},
\end{equation}
and according to Eq. (\ref{E5})
\begin{equation}
\label{E15}
\phi'(k_x,k_y) = \mathcal{F}\{\psi'(x,y,z_0)\}e^{-i\sqrt{k^2-k_x^2-k_y^2}z_0}.
\end{equation}
Thus after the aperture, $\vec{p}_{eff}$ and $\vec{k}_{eff}$ were changed according to Eq. (\ref{E7}) and Eq. (\ref{E8}).

\textit{Local photon momentum.}---The effective photon momentum and effective wave vector in Eq. (\ref{E7}) and Eq. (\ref{E8})  are for the whole field and nonlocal. Due to the uncertainty principle, the determination of photon position, realized by interaction or measurement, could at the same time affect $\vec{k}_{eff}$. For example, if we provide a transverse confinement of photon position with $t(x,y) = \delta(x-x_0,y-y_0)$ on plane $z_0$, according to Eq. (\ref{E14}), in this case
\begin{equation}
\label{E16}
\psi'(x,y,z_0) = \delta(x-x_0,y-y_0).
\end{equation}
After taking Eq. (\ref{E16}) to Eq. (\ref{E15}) we have 
\begin{equation}
\label{E17}
\phi'(k_x,k_y) = e^{-i(k_xx_0+k_yy_0)}e^{-i\sqrt{k^2-k_x^2-k_y^2}z_0}.
\end{equation}
Taking Eq. (\ref{E17}) into Eq. (\ref{E8}), we get the effective wave vector at the position $(x_0,y_0,z_0)$ as
\begin{equation}
\label{E18}
\vec{k}_{eff} = \frac{2}{3}k\vec{e}_{k_z},
\end{equation}
and the effective photon momentum as
\begin{equation}
\label{E19}
\vec{p}_{eff} = \frac{2}{3}\hbar k\vec{e}_{k_z}.
\end{equation}
The result is different with the canonical photon momentum defined in \cite{RN11} and local wave vector in \cite{RN2} that described by the optical phase gradient at the given position.

\textit{Transverse confinement in light atom interaction.}---We analysis the interaction between a structured field $u(x,y,z)e^{-i\omega t}$ and a single two-level atom. In the dipole approximation, the interaction Hamiltonian is \cite{RN18} 
\begin{equation}
H_{de} = -\vec{D}\cdot\vec{E}(x_0,y_0,z_0).
\end{equation}
$(x_0,y_0,z_0)$ is the center position of the atom, $\vec{E}$ is electric field and $\vec{D}$ is the electric dipole. The atom is supposed to be located at the given position and interact with the local field through dipole interaction. 

In real case, the atom position distribution is described by its wave function $\psi_{atom}(x,y,z)$. When the atom absorbs one photon on plane $z_0$, the atom transverse probability distribution could also provide a confinement for the absorbed photon. Comparing to the aperture, which projects the transverse photon state to $\ket{x,y}$ with probability $t(x,y)$, we define the projection operator for the atom on plane $z_0$ as
\begin{equation}
\hat{O} = \iint |\psi_{atom}(x,y,z_0)|^2\ket{x,y}\bra{x,y}dxdy.
\end{equation}
When an atom absorbs one photon on this plane, the photon transverse state becomes
\begin{equation}
\ket{photon'} = \hat{O}\ket{photon}.
\end{equation}
According to Eq. (\ref{E6}),
\begin{equation}
\label{E23}
\begin{aligned}
\psi'(x,y,z_0) =\frac{\psi(x,y,z_0)|\psi_{atom}(x,y,z_0)|^2}{\sqrt{\iint |\psi(x,y,z_0)|\psi_{atom}(x,y,z_0)|^2|^2dxdy  }},
\end{aligned}
\end{equation}
and
\begin{equation}
\label{E24}
\phi'(k_x,k_y) = \mathcal{F}\{\psi'(x,y,z_0)\}e^{-i\sqrt{k^2-k_x^2-k_y^2}z_0}.
\end{equation}
Taking Eq. (\ref{E24}) to Eq. (\ref{E8}), we can get $\vec{k}_{eff}$ for the absorbed photon.

Fig \ref{Fig.1}.(c) shows the interaction between a plane wave light and a Gaussian atom cloud with radius $\sigma_a$. In this case, $\psi(x,y,z)=e^{ikz}$ and the atom transverse probability distribution is independent on the longitudinal coordinate $z$
\begin{equation}
\label{E25}
|\psi_{atom}(x,y)|^2 = \frac{1}{2\pi\sigma_a^2}e^{-\frac{x^2+y^2}{2\pi\sigma_a^2}}.
\end{equation}
According to Eq. (\ref{E23}) and Eq. (\ref{E24}) we have $|\phi'(k_x,k_y)|^2=(1/\pi\sigma_a^2)e^{-\sigma_a^2(k_x^2+k_y^2)}$. Then $\vec{k}_{eff}$ is dependent on $\sigma_a$.

\begin{figure}[h]
	\centering
	\includegraphics[width=8.6cm]{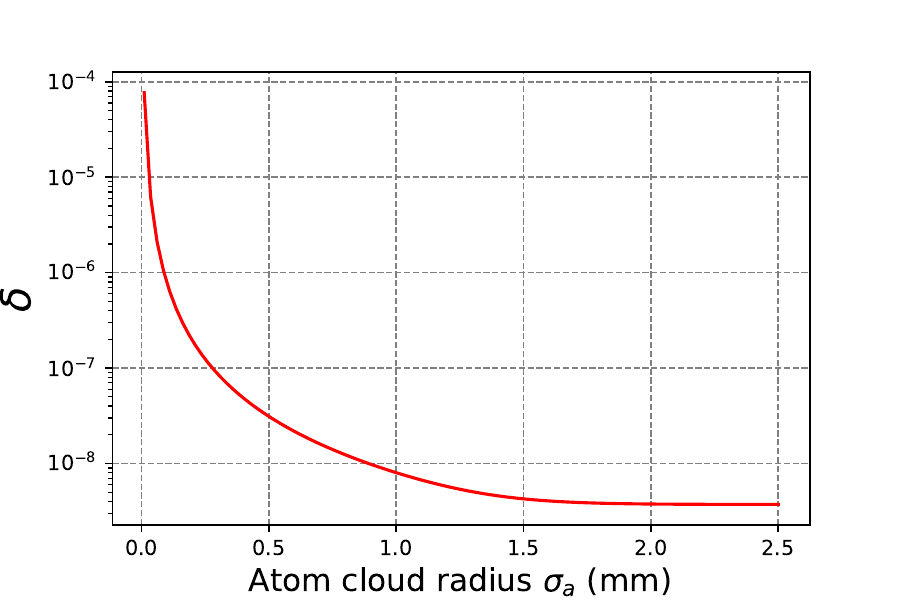}
	\caption{$\sigma_B = 3.6$ mm, beam wavelength $\lambda = 780$ nm, $\delta = (k-k_{eff})/k$ as a function of atom cloud radius.}
	\label{Fig.2}
\end{figure}
In experiment we usually use Gaussian beams. Suppose the beam waist lies on $z=0$ plane, here the photon transverse position probability amplitude is
\begin{equation}
\label{E26}
\psi(x,y,0)=\frac{1}{\sqrt{2\pi}\sigma_B}e^{-\frac{(x^2+y^2)}{4\sigma_B^2}},
\end{equation}
$\sigma_B$ is the beam waist. When an atom in the cloud of Eq. (\ref{E25}) absorbs one photon on $z=0$ plane, $\vec{k}_{eff}$ depends on both the atom cloud radius $\sigma_a$ and beam waist $\sigma_B$. In this case $k_{x,eff} = k_{y,eff} = 0$, and $\vec{k}_{eff}=k_{z,eff}\vec{e}_{k_z}$. We define $\delta$ as the relative shift of $k_{eff}$ to $k$,
\begin{equation}
\delta = \frac{k-k_{eff}}{k}.
\end{equation}

In Fig. \ref{Fig.2}, we have plotted the value of $\delta$ as a function of $\sigma_a$. Setting the wavelength $780$ nm and the beam waist $3.6$ mm, $\delta$ changes from the order of $10^{-9}$ to $10^{-5}$ as the atom cloud radius decreases from $2.0$ mm to $10$ um. Depending on the atom temperature, those are normal sizes for optical molasses and Bose-Einstein condensation (BEC). A smaller atom cloud radius could lead to a larger relative shift of $k$.
\begin{figure}[h]
	\centering
	\includegraphics[width=8.6cm]{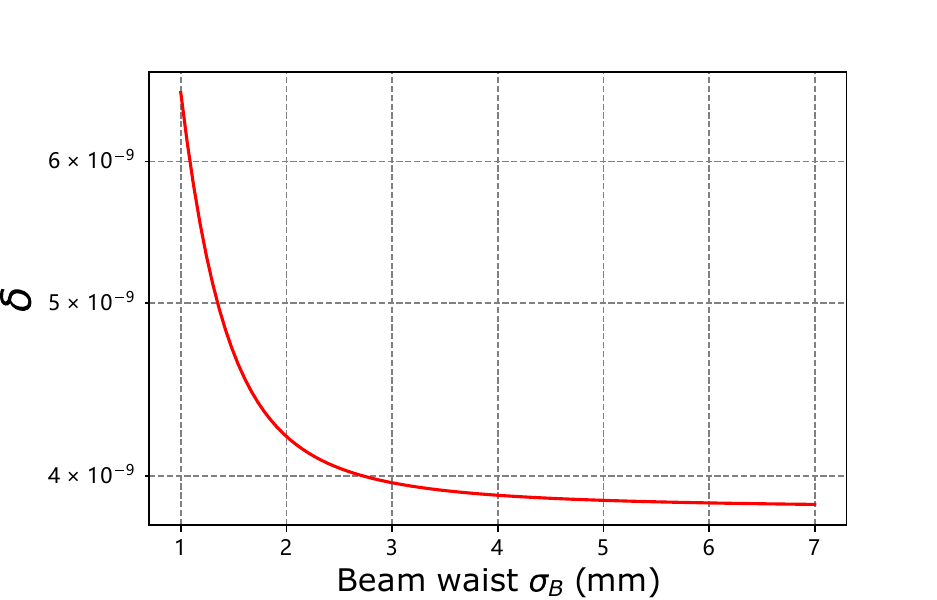}
	\caption{$\sigma_a = 1.7$ mm, beam wavelength $\lambda = 780$ nm,  $\delta = (k-k_{eff})/k$ as a function of beam waist.}
	\label{Fig.3}
\end{figure} 

In Fig. \ref{Fig.3}, we keep the atom radius as $1.7$ mm and change the beam waist from $1$ mm to $7$ mm, there is a decrement of $\delta$ at the order of $10^{-9}$. Similarly, a smaller beam waist could lead to a larger relative shift of $k$.

At one extreme case, if both the atom matter wave and the light are in plane wave mode, as shown in Fig. \ref{Fig.1}.(b), there will be no confinement and $\delta = 0$. On the other extreme, by setting $\sigma_a$ the value of Bohr radius, we get $\delta \approx 0.33 $. We define the atom mean recoil velocity as
\begin{equation}
\vec{v}_r=\frac{\hbar \vec{k}_{eff}}{m},
\end{equation}
$m$ is the atom mass. As in our model $k_{eff}$ is always smaller than $k$, there will be no recoil velocity larger than $\hbar k/m$.

\textit{Systematic effect in atom interferometer.}---In light-pulsed atom interferometer, the total phase consists of two parts. One part depends on laser frequency and the other part depends on atoms' motion \cite{RN14,RN19}. The phase accumulated by atoms relies on $\vec{k}_{eff}$ of the absorbed photons \cite{RN10}. The relative shift of wave vector could lead to a shift of the final result in high precision measurements. For the measurement of fine structure constant $\alpha$, the result directly relies on the measurement of the mean recoil velocity $\vec{v}_r$. 

In the $\alpha$ measurement  \cite{RN12,RN20}, the principle is to transfer an amount of recoil velocities to an atom cloud using a 1D accelerating lattice, then use a Ramsey-Bord\'e interferometer to measure the mean recoil velocity. The relative correction of $\alpha$, $\Delta\alpha/\alpha$ scales as the relative shift of wave number $\Delta k/k = -\delta$ \cite{RN12}, also the case is for the relative correction of recoil velocity $v_r$. Here $\Delta k = k_{eff}-k$.

From Eq. (\ref{E23}), Eq. (\ref{E24}) and Eq. (\ref{E8}) we can see that in our model $\vec{k}_{eff}$ and $\delta$ depend on the atom wave function and the optical transverse complex amplitude. 

In two recent works \cite{RN14,RN11}, the experimenters found a relative shift of recoil velocity $v_r$ as they changing the relative number of atoms. In their experiment they change the number of atoms by changing the intensity of the lattice laser thus to change the efficiency of Bloch oscillation \cite{RN21}. In fact at the same time the spatial distribution of the survival atoms is also changed.

In \cite{RN14} with lower lattice intensity, fewer atoms survived at high intensity region, the center of the Gaussian beam, a negative relative shift to the level $10^{-8}$ of $v_r$ has been observed. That could correspond to a decrease of atom cloud radius by several $0.1$ mms from around $0.5$ mm in Fig. (\ref{Fig.2}) in our model. By blowing away atoms in the high intensity region of the lattice laser, a positive shift of the relative value of $v_r$ has been observed at the magnitude of $10^{-8}$, and it was regarded as an extra photon recoil \cite{RN11}. It could also be explained with our model to some extent. Because by doing that the center part of the atom cloud was lost, located at the center of the Gaussian beam, and the residual atoms has a more incompact spatial distribution. Then according to Eq. (\ref{E24}), $\psi'(k_x,k_y)$ could be more compact. That could lead to a smaller $\delta$ in principle. In our opinion the measured recoil velocity is not an extra recoil, but a relative larger recoil velocity compared to a previous value \cite{RN12}.

\textit{Summary}---Based on plane wave angular spectrum, we have analyzed the nonlocal effective wave vector of photon in structured light field. We proposed that in light-atom interaction, the atom transverse wave function could provide a 2D confinement to photons. As a result, the effective wave vector of the absorbed photons and the mean recoil velocity could be affected. 

We have defined a projection operator to describe the confinement of photon transverse position state, and calculated the relative shift of the effective wave vector when an atom with a Gaussian wave function absorbs one photon at the waist of a Gaussian beam. This shift could lead to a systematic effect related to atom cloud size and Gaussian beam waist in high precision measurement based on light-pulsed atom interferometer. Our model complies with uncertainty principle and sheds new light on momentum transfer in light-atom interaction.  

\textit{Acknowledgement.}---This work was supported by China Scholarship Council(CSC) and Foundation Franco-Chinoise pour la Science et ses Applications(FFCSA).

\bibliographystyle{apsrev4-1}  
\bibliographystyle{unsrt}
\bibliography{ref}
\end{document}